\documentclass[%
reprint,
 amsmath,amssymb,
 aps,
]{revtex4-2}
\usepackage{titlesec}
\titlespacing*{\subsection}{0pt}{6pt}{2pt}
\usepackage{hyperref}
\usepackage{float}
\usepackage{nicematrix}
\usepackage{graphicx}
\usepackage{dcolumn}
\usepackage{bm}

\begin{document}

\preprint{APS/123-QED}

\title{Tensor-Based Chaotic Perception}

\author{Amir M. Majd}
\email{amir.mahmoudimajd@gmail.com}

\date{\today}

\begin{abstract}
Symbols can provide compact representations of complex sensory information. Mathematically, chaotic attractors can provide a basis for such representations, with object identity encoded in the topological organization of the attractor. In three-dimensional phase space, this organization can be characterized by the linking matrix of unstable periodic orbits, which serves as a topological fingerprint. Because the linking matrix has no closed-form dependence on the learnable parameters, we learn the chaotic series generating the target attractors rather than optimizing the invariant directly; the reconstructed attractor then determines its unstable periodic orbits and linking matrix through its topology. In this paper, we propose a third-order tensor that maps inputs to connection matrices that generate the corresponding chaotic series. Exploiting the duality between a vector and its sequential representation as a series, we lift inputs into a higher-dimensional space and reorder their components, allowing the input to be presented sequentially rather than simultaneously. We first demonstrate that this representation can discriminate between classes in the \href{https://archive.ics.uci.edu/dataset/151/connectionist+bench+sonar+mines+vs+rocks}{Sonar} dataset. We then examine perceptual constancy using the \href{https://www.csc.kth.se/cvap/databases/kth-tips/index.html}{KTH-TIPS2-a} dataset, where variations in color and texture within a class are mapped to distinct chaotic series associated with the same topological attractor class. The results demonstrate that the proposed tensor-based mapping can preserve attractor class while accommodating variations in the sensory realization of an object.
\end{abstract}

\keywords{Randomly connected network, chaos, linking numbers, Learning}         
\maketitle
\raggedbottom  

\section{Symbolic Perception}
\begin{figure}
\begin{center}
\includegraphics[width=5cm]{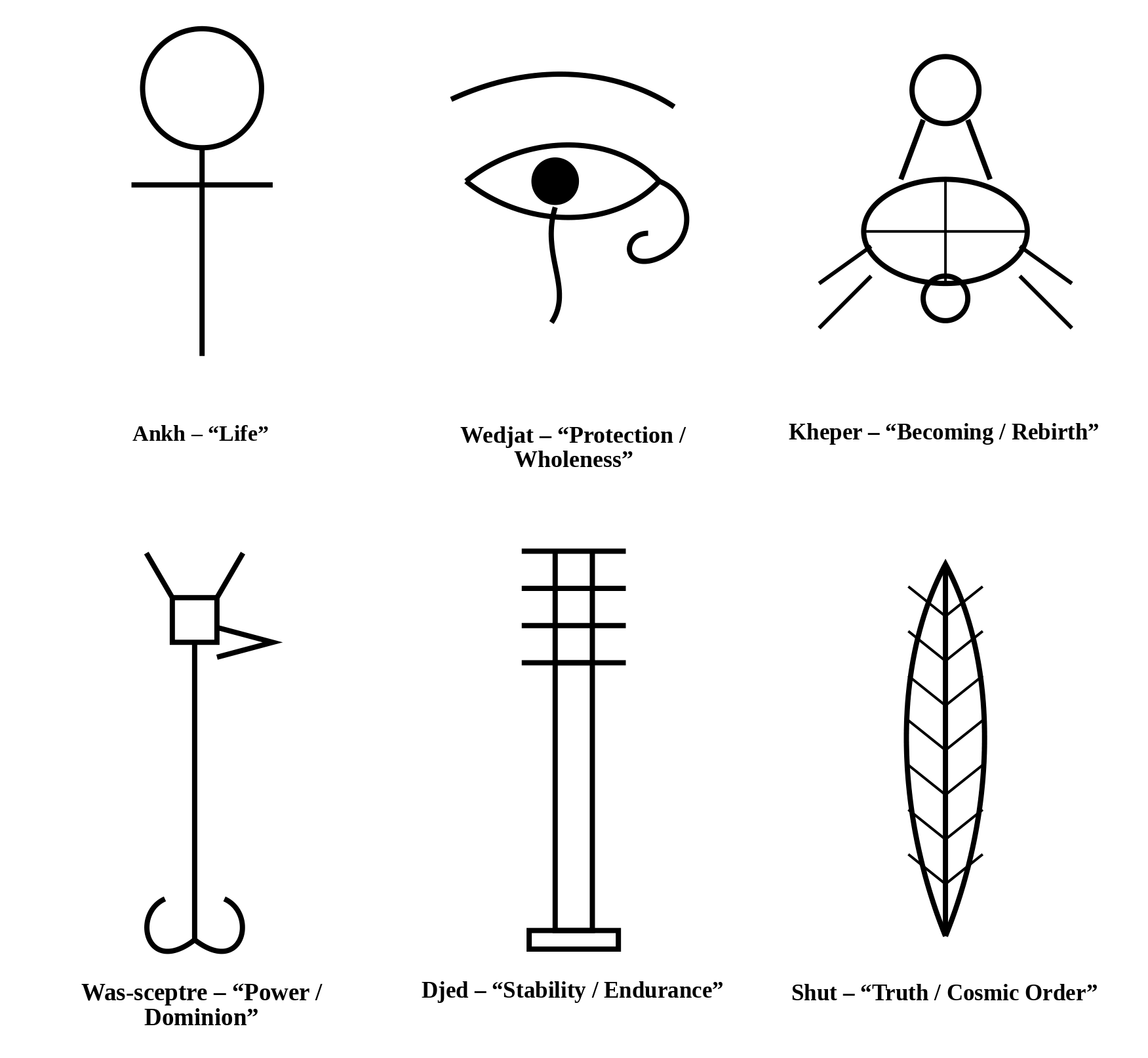}
\caption{Examples of symbolic representations that condense complex semantic content into compact visual forms.}
\label{fig:symbols}
\end{center}
\end{figure}

Human cognition is distinguished by its capacity for symbolic representation. Perception, memory, and communication rely on transforming complex experiences into compact representations that can be stored, manipulated, and shared. Human symbolic processing is qualitatively different from that of earlier hominids \cite{Tattersall2010}, and contemporary accounts propose that symbolic systems discretize complex domains into small sets of symbols that can be combined into structured representations, effectively encoding and compressing information \cite{Dehaene2022}. Archaeological evidence likewise suggests that symbolic thought became a defining feature of modern human cognition, with early symbolic behavior expressed through art and other material representations \cite{Tattersall2010,Miyagawa2018}. Across cultures and historical periods, humans have transformed sensory and conceptual information into stable, reusable symbolic forms, from cave paintings and religious iconography to mathematical notation and traffic signs. Symbols can therefore be viewed as cognitive condensations of experience, reducing complex semantic structures to compact and identifiable representations. Ancient Egyptian hieroglyphs illustrate this principle: glyphs such as the Ankh, Djed, and Was-sceptre as depicted in Fig.~\ref{fig:symbols} compressed broad conceptual domains into visually compact forms. Their historical specificity is less important here than the general cognitive strategy they exemplify: reducing complex semantic structures to identifiable symbolic signatures.

Semiotics provides a formal framework for understanding such symbolic representation. In the tradition of Peirce, a symbol forms part of a triadic relation involving a sign, an object, and an interpretant, with meaning emerging through their interaction rather than through fixed correspondences. Modern cognitive science similarly emphasizes that meaning is grounded in embodied and sensorimotor experience. Perception is inherently multimodal and context dependent, and symbolic representations can emerge through recurrent interactions between an organism and its environment. Symbols are therefore not simply imposed upon perception; they can arise from perceptual processes themselves. From this perspective, perception may be viewed as the construction of internal invariants that remain stable despite changes in sensory conditions. The central question motivating the present work is whether an analogous mechanism exists at the neural level: if perception produces stable internal representations of objects, how can a neural system generate representations that preserve object identity despite sensory variability? The present framework proposes that such invariants are not merely geometric or statistical, but topological.

\section{Chaos, Memory, and Perception}
A seminal contribution was made by Skarda and Freeman \cite{Skarda1987}, who proposed that chaos constitutes the ground state of the perceptual system. In their framework, learned stimuli correspond to attractors embedded within a chaotic background state. Chaos allows rapid escape from existing attractors, facilitating novelty detection while preserving the capacity to form new memories. Subsequent experimental and theoretical studies strengthened this view. Faure and Korn \cite{Faure2001,Korn2003} reviewed evidence for chaotic dynamics across multiple scales of neural organization, from individual neurons to large-scale cortical networks. Tsuda's theory of chaotic itinerancy \cite{Tsuda1991,Tsuda2015} further proposed that cognition arises through transitions among quasi-attractors representing stored memories, providing a mechanism for flexible learning without catastrophic interference. Related ideas were supported by electrophysiological observations and by models of dynamic associative memory operating near the boundary between order and chaos.

Together, these studies associate chaos with several functions relevant to perception and learning, including stimulus discrimination, novelty detection, attentional flexibility, and memory retrieval \cite{Skarda1987,Tsuda2001,Tsuda2015}. More recent work has extended these ideas to pattern recognition and reservoir computing, where complex recurrent dynamics are exploited computationally rather than suppressed.

Despite these advances, existing approaches remain primarily phase-space descriptions. They characterize perception in terms of attractors and trajectories but do not address whether the topology of an attractor itself carries representational content. Consequently, they leave unresolved the problem of perceptual invariance: how an object can retain a stable identity despite substantial variation in its sensory realization. The present framework addresses this gap by shifting attention from the dynamics of attractors to their topology.

\section{The Central Hypothesis and the implementation challenges}
Before presenting the central hypothesis, let's review linking matrices first: 
\subsection{Linking matrix}
Strange attractors are characterized by two complementary classes of attributes: dynamical and topological. The dynamical attributes, often regarded as the traditional quantifiers of chaos, include the fractal dimension, metric entropy, and the spectrum of Lyapunov exponents \cite{Hilborn1994}. The topological attributes, which serve as invariant ``fingerprints'' of strange attractors, include linking numbers, topological entropy, and rotation rates (for phase spaces of the form \(\mathbb{R}^{3}\times S^{1}\)). Because topological organization becomes increasingly difficult to define in higher-dimensional phase spaces, these topological fingerprints are primarily applicable to attractors embedded in three-dimensional space \cite{Gilmore2011}.A chaotic trajectory repeatedly approaches neighborhoods of unstable periodic orbits (UPOs). Segments of the trajectory that closely shadow a periodic orbit are commonly referred to as \emph{surrogate periodic orbits}; upon stabilization, these yield the corresponding UPOs. From this perspective, the dynamics on a strange attractor may be viewed as a sequence of transitions, or ``hopping,'' among different UPOs.

\begin{figure}
\begin{center}
\includegraphics[width=3.5cm]{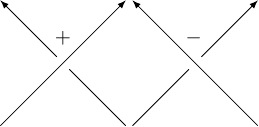}
\caption{Crossing-sign convention used in the computation of linking numbers. The assigned signs are invariant under clockwise or counterclockwise rotations of the projection plane.}
\label{fig:linkconvention}
\end{center}
\end{figure}

The topological organization of an attractor is encoded in its linking matrix. To construct this matrix, one first extracts a collection of UPOs from the chaotic attractor and projects them onto a common plane. Crossings between pairs of UPOs are then identified and assigned signs according to the convention illustrated in Fig.~\ref{fig:linkconvention}. 

The linking number between two distinct UPOs \(\mathcal{U}_1\) and \(\mathcal{U}_2\) is defined as
\begin{equation}
L(\mathcal{U}_1,\mathcal{U}_2)
=
\frac{1}{2}
\sum_i
\alpha_i(\mathcal{U}_1,\mathcal{U}_2),
\end{equation}
where \(\alpha_i=\pm1\) denotes the sign of the \(i\)-th crossing. The self-linking number is similarly defined by
\begin{equation}
L(\mathcal{U}_1,\mathcal{U}_1)
=
\sum_i
\alpha_i(\mathcal{U}_1,\mathcal{U}_1).
\end{equation}
The collection of pairwise linking numbers forms the \emph{linking matrix}, which serves as a topological fingerprint of the attractor. The off-diagonal entries record the linking numbers between distinct UPOs, while the diagonal entries correspond to their self-linking numbers. Attractors belonging to the same topological class possess identical off-diagonal linking structures, making the linking matrix a powerful tool for distinguishing topologically inequivalent chaotic attractors.

\subsection{The central hypothesis}
\emph{Object identity is encoded in the topological class of a chaotic attractor generated by a recurrent neural network when driven by the object's sensory representation. The invariant of this class---the linking matrix of the unstable periodic orbits embedded within the attractor---constitutes the object's mathematical fingerprint. Learning establishes this fingerprint, while recognition reconstructs it from sensory input.}
\subsection{The implementation challenges}
We begin by addressing two fundamental questions: Which neuronal model is most suitable for chaos-based perception? What is responsible for the generation of attractors? 

One neuronal model generating chaos is the recurrent continuous-time neural network (RCNN) of the form\cite{Sompolinsky1988}
\begin{equation}
\dot{X}_{i}(t)=-X_{i}(t)+g\sum_{j=1}^{N}J_{ij}\phi(X_{j}(t)),
\end{equation}
where $X_i(t)$ denotes the activity of neuron $i$, $g$ is a gain parameter controlling interaction strength, and $\phi(X)=\tanh(X)$ is a sigmoidal activation function mapping membrane potentials to firing rates. The connectivity matrix $\mathbf{J}$ is drawn from a Gaussian ensemble with zero mean, variance $1/N$, and correlation $[J_{ij},J_{ji}]=\sigma/N$, with $\sigma\in[-1,1]$. 

\begin{figure}
\begin{center}
\includegraphics[width=8cm]{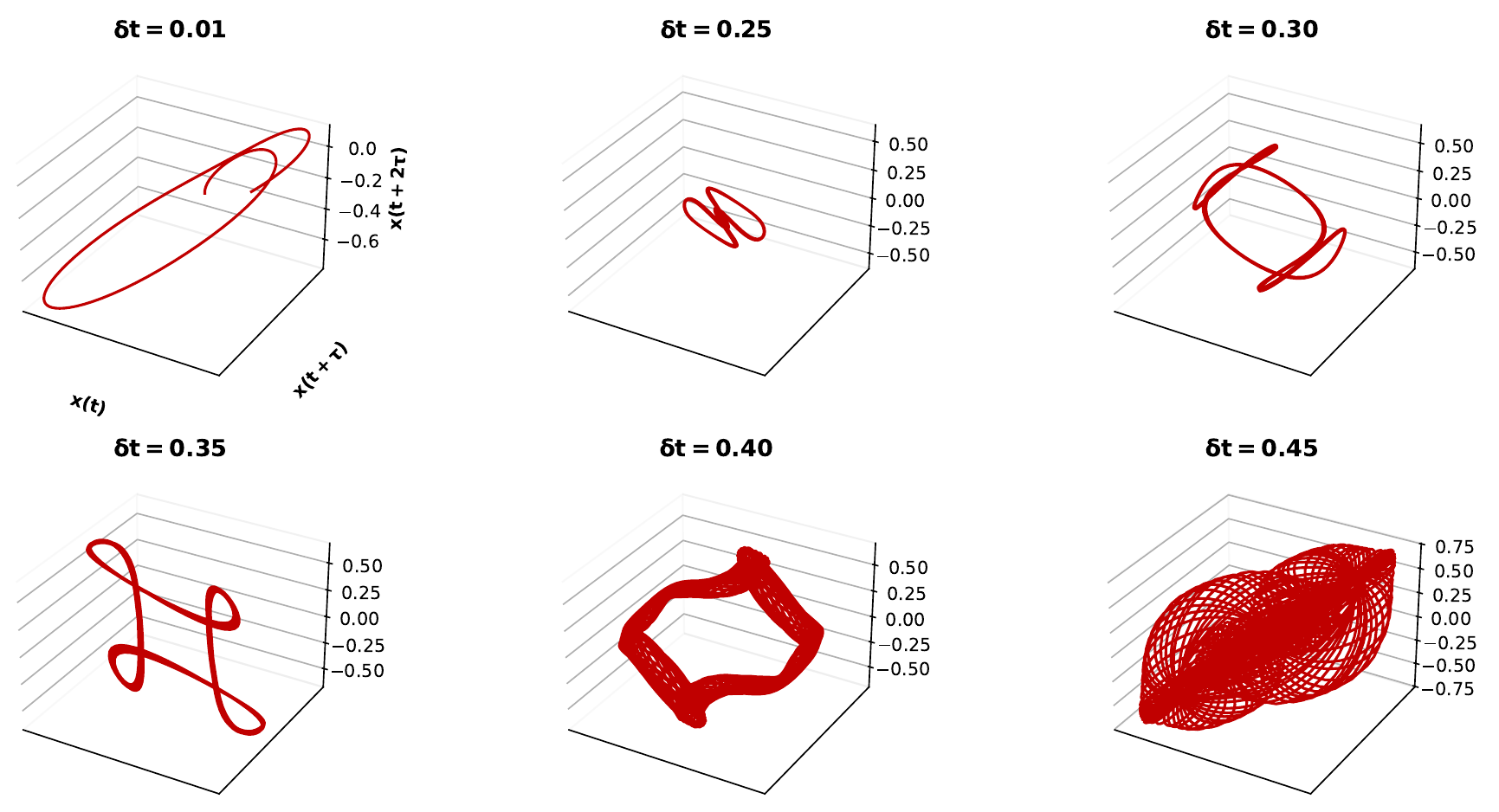}
\includegraphics[width=8cm]{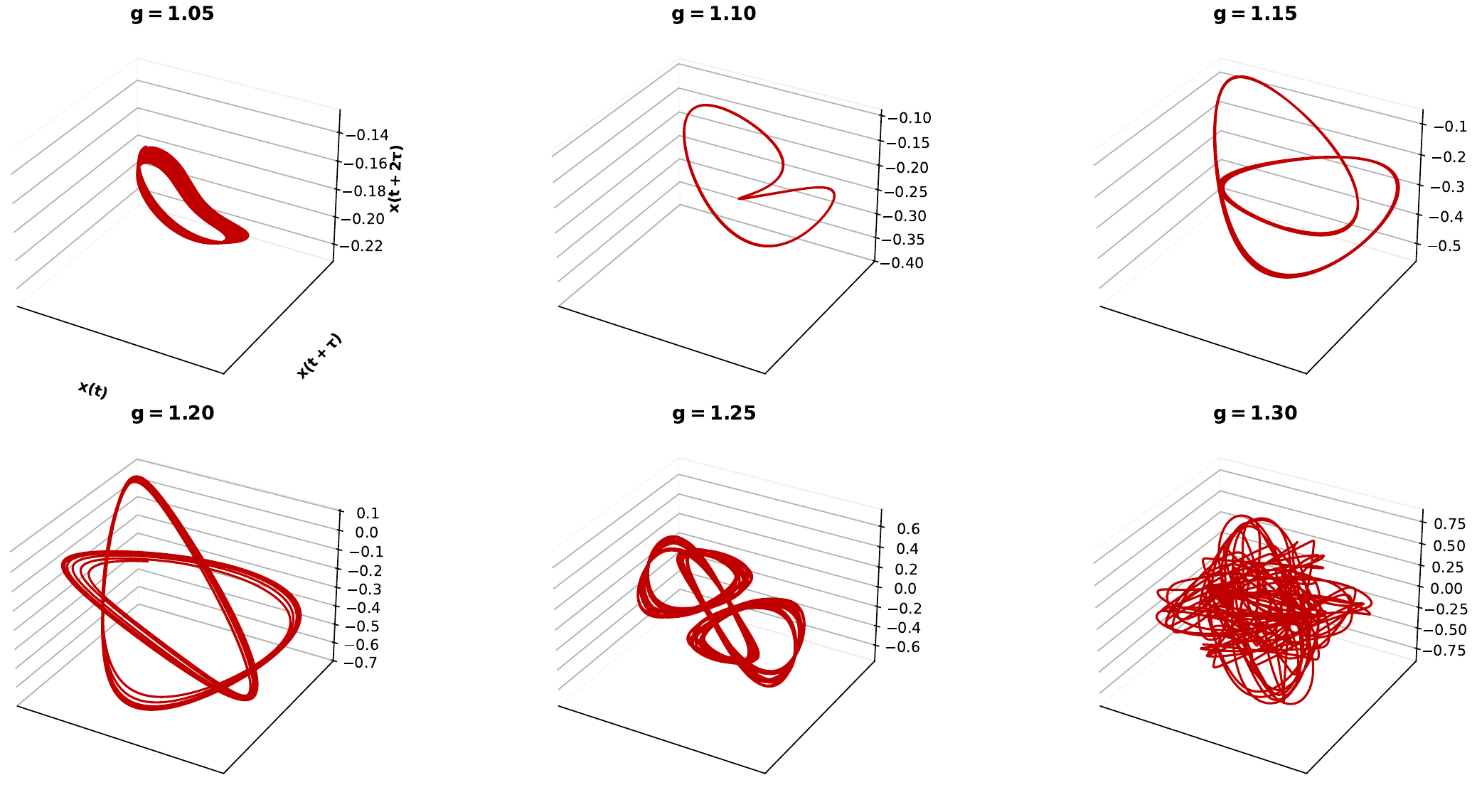}
\includegraphics[width=8cm]{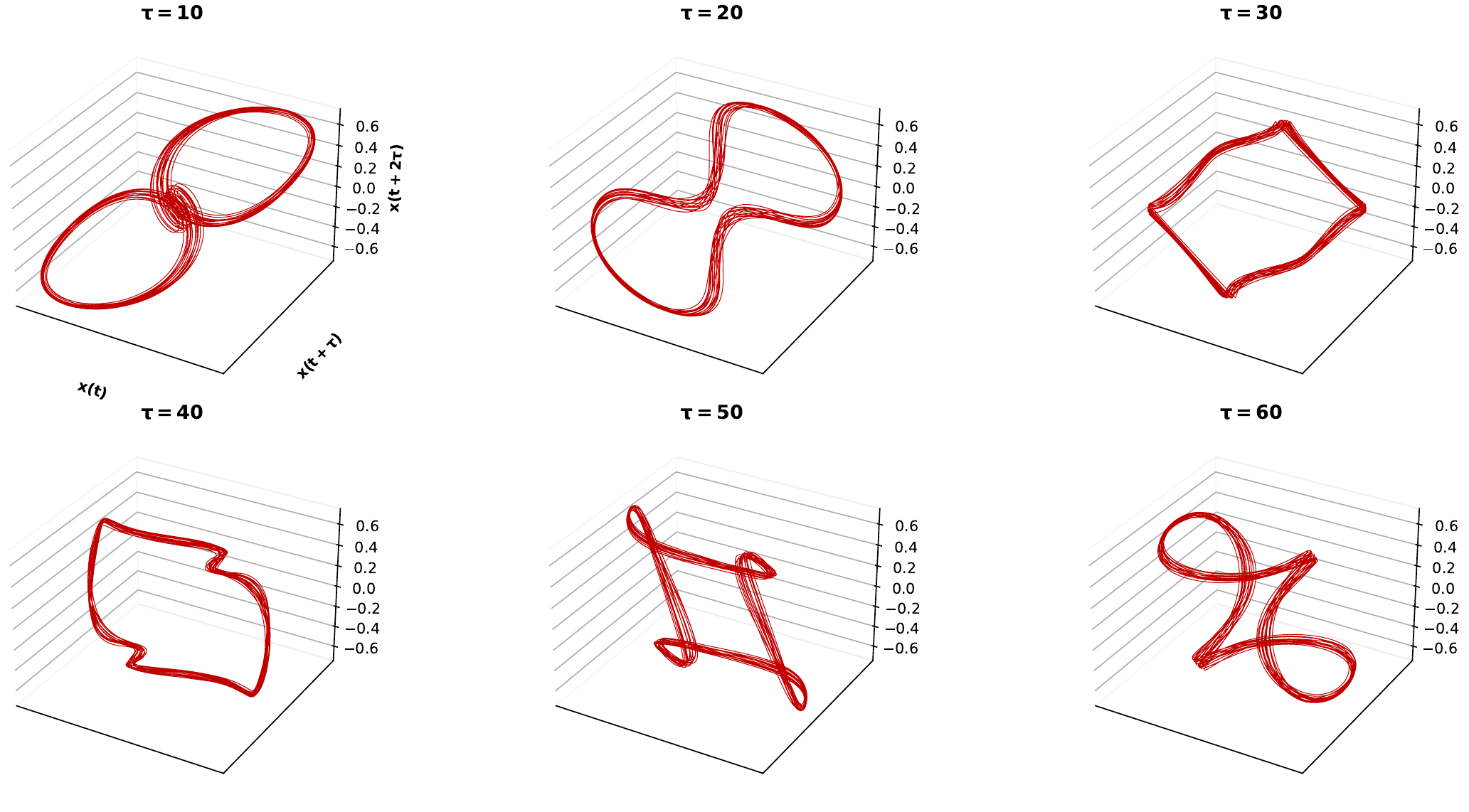}
\caption{The sensitivity of the network's reconstructed attractors (from series of length 5000) to time steps ($g=1.22,\tau=85$), to g ($dt=0.25,\tau=65$), and time lag ($dt=0.2,g=1.25$).}
\label{fig:tau}
\end{center}
\end{figure}

No matter whether a single neuron or a group of neurons generates the sought chaotic dynamics, implementing this model faces several fundamental challenges. First, perception is essentially instantaneous: once the sensory input is received, perception is completed immediately. In the proposed model, however, one must wait for N time steps—that is, the neurons must evolve for N iterations before an attractor containing sufficient information about its topological organization can emerge. Second, as Fig.~\ref{fig:tau} shows, both the integration time step and the gain parameter g have a substantial impact on the resulting attractor. Different choices of the time delay used for attractor reconstruction also lead to different topological organizations. The situation is further complicated by the limited efficiency of current algorithms for extracting unstable periodic orbits and by the lack of a closed-form method for computing linking matrices. As a result, one is forced to rely on heuristic optimization techniques, such as genetic algorithms, making the search for the desired topological structure akin to finding a needle in a haystack.

\section{From Temporal Chaos to Spatial Representation; the vector-series duality}
We adopt a different perspective from conventional approaches. Rather than interpreting chaos as a temporal phenomenon, we regard the instantaneous population state
\begin{equation}
\mathbf{X}\in\mathbb{R}^{N}
\end{equation}
as a spatial representation of a chaotic object. The components of $\mathbf{X}$ are interpreted as ordered samples of a global structure encoded collectively by the neuronal population. The objective is therefore not to reproduce the temporal evolution of an attractor, but to reproduce its topology within a single population state. This viewpoint makes the construction of an explicit learning objective substantially simpler.

Inspired by the discretized form of Eq.~(3),
\begin{equation}
\mathbf{X}(t_{0}+\delta)
=
(1-\delta)\mathbf{X}(t_{0})
+
\delta g
\mathbf{J}
\phi\left(\mathbf{X}(t_{0})\right),
\label{eq:model_discrete}
\end{equation}
setting $\delta=1$, absorbing $g$ into $\boldsymbol{J}$, and omitting the activation function gives
\begin{equation}
\mathbf{X}(t_{0}+1)
=
\mathbf{J}
\mathbf{X}(t_{0}),
\label{eq:model_simple}
\end{equation}
where $\boldsymbol{J}$ is an $N\times N$ matrix that maps the population state $\mathbf{X}(t_{0})$ to the next state $\mathbf{X}(t_{0}+1)$ in a single time step.

The interpretation of $\mathbf{X}$ as an attractor introduces a {\em vector--series duality} when read vertically. Thus, the attractor can be considered either as a point in an $N$-dimensional space or, through the ordering of its components, as a series. This duality provides a useful means of reorganizing information without changing its underlying values and becomes particularly relevant when distinguishing low-dimensional samples belonging to closely related classes.

\subsubsection*{The learning rule}
Let the sensory input be represented by $\boldsymbol{Y}^{\alpha}$, $\alpha$ indicating the class. Then, the synaptic weight matrix is generated adaptively from the encoded activity through a third-order tensor,
\begin{equation}
J_{ij}^{(\alpha)}
=
\sum_{k=1}^{N}
W_{ijk}^{(\alpha)}
Y_k^{(\alpha)}+C_{ij}^{(\alpha)}
\end{equation}
The corresponding network output is
\begin{equation}
O_i^{(\alpha)}
=
\sum_{j=1}^{N}
J_{ij}^{(\alpha)}
Y_j^{(\alpha)},
\end{equation}
which may be written explicitly as
\begin{equation}
O_i^{(\alpha)}
=
\sum_{j=1}^{N}
\sum_{k=1}^{N}
W_{ijk}^{(\alpha)}
Y_j^{(\alpha)}
Y_k^{(\alpha)} .
\end{equation}
where we put $C_{ij}=0$ for simplicity. 
The loss is defined as 
\begin{equation}
\mathcal{L}(W^{\alpha})
=
\frac12
\sum_{i=1}^{N}
\left(
O_i^{(\alpha)}
-
A_i^{(\alpha)}
\right)^2,
\end{equation}
Defining the reconstruction error
\begin{equation}
E_i^{(\alpha)}
=
O_i^{(\alpha)}
-
A_i^{(\alpha)},
\end{equation}
the loss assumes the compact form
\begin{equation}
\mathcal{L}
=
\frac12
\sum_i
\left(
E_i^{(\alpha)}
\right)^2.
\end{equation}
Differentiating the network output with respect to the tensor elements yields
\begin{equation}
\frac{\partial O_i^{(\alpha)}}
{\partial W_{abc}^{(\alpha)}}
=
\delta_{ia}
Y_b^{(\alpha)}
Y_c^{(\alpha)},
\end{equation}
where $\delta_{ia}$ denotes the Kronecker delta.
Applying the chain rule,
\begin{equation}
\frac{\partial\mathcal{L}}
{\partial W_{abc}^{(\alpha)}}
=
\sum_i
E_i^{(\alpha)}
\frac{\partial O_i^{(\alpha)}}
{\partial W_{abc}^{(\alpha)}},
\end{equation}
gives
\begin{equation}
\frac{\partial\mathcal{L}}
{\partial W_{abc}^{(\alpha)}}
=
E_a^{(\alpha)}
Y_b^{(\alpha)}
Y_c^{(\alpha)},
\end{equation}
or, equivalently,
\begin{equation}
\frac{\partial\mathcal{L}}
{\partial W_{ijk}^{(\alpha)}}
=
\left(
O_i^{(\alpha)}
-
A_i^{(\alpha)}
\right)
Y_j^{(\alpha)}
Y_k^{(\alpha)}.
\end{equation}

\subsection*{Sonar dataset}
\begin{figure}
\begin{center}
\includegraphics[width=7.5cm]{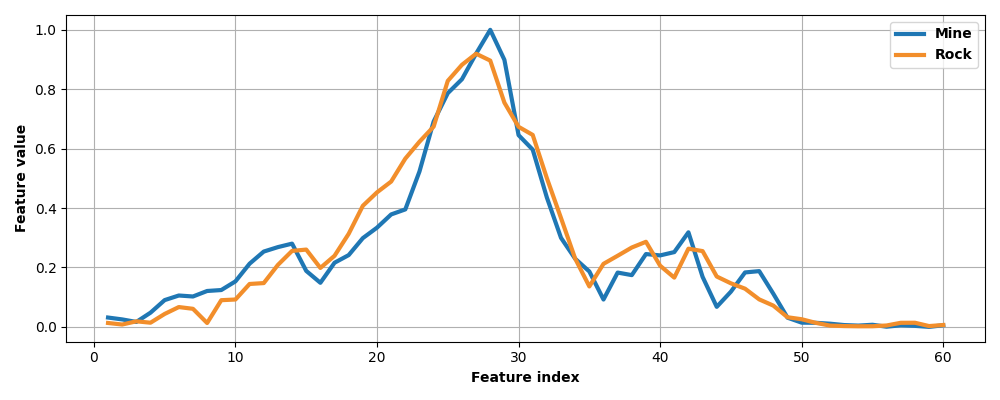}
\includegraphics[width=7.5cm]{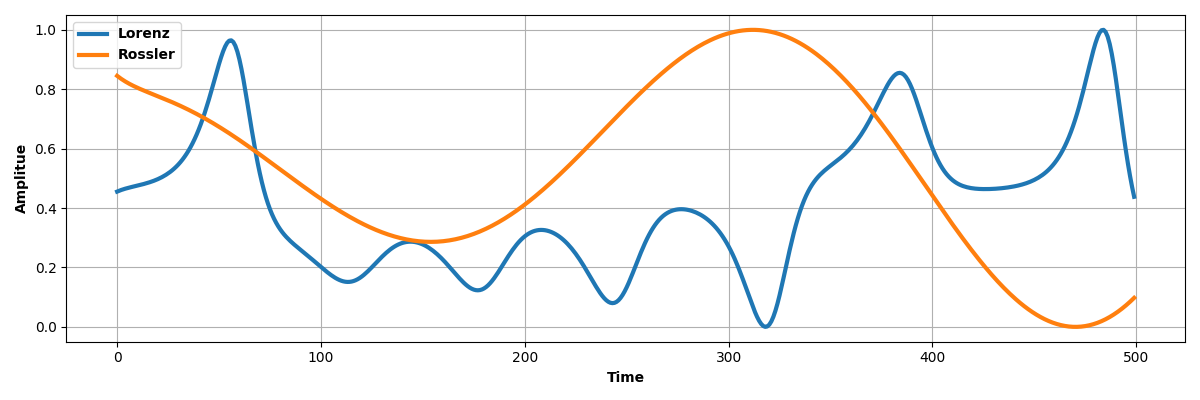}
\caption{Closest pair of feature vectors from the Sonar dataset, with Euclidean distance 0.5073 between one Rock sample and one Mine sample (up), and the corresponding Lorenz and R\"{o}ssler target series assigned to the Rock and Mine classes, respectively (down).}
\label{fig:sonar_proximity}
\end{center}
\end{figure}

The Sonar dataset\cite{gorman1988sonar}, where mine and rock classes are associated with the $x$ of Lorenz 
\begin{equation}
\begin{aligned}
\dot{x} &= \sigma(y-x)\\
\dot{y} &= x(\rho-z)-y\\
\dot{z} &= xy-\beta z
\end{aligned}
\end{equation}
($\sigma=10,\rho =28,\beta=8/3$) and $x$ of R\"{o}ssler
\begin{equation}
\begin{aligned}
\dot{x} &= -y-z\\
\dot{y} &= x+a y\\
\dot{z} &= b+z(x-c)
\end{aligned}
\end{equation}
($a=0.2,b=0.2,c=5.7$) respectively (Fig.~\ref{fig:sonar_proximity}, up), provides a demanding classification problem because samples belonging to different classes can remain closely proximate in the original feature space. Fig.~\ref{fig:sonar_proximity}(down) illustrates this proximity. Each sample is represented by a $60\times1$ feature vector, which we embed into the $N$-dimensional neuronal space by padding it with zeros. However, applying the same zero-padding scheme to both classes does not resolve the proximity problem: two nearby feature vectors remain nearby after embedding.

The vector--series duality offers a simple way around this limitation. Since the components of $\mathbf{X}$ can be interpreted as an ordered series, the zero-padding need not be identical for the two classes. For a mine, for example, the 60 features may occupy the first 60 positions,
\begin{equation}
\mathbf{Y}^{(M)}
=
(x_1,\ldots,x_{60},0,\ldots,0)_{1\times500}^{\mathsf T},
\end{equation}
whereas for a rock they may be shifted to the next 60 positions,
\begin{equation}
\mathbf{Y}^{(R)}
=
(0,\ldots,0,x_1,\ldots,x_{60},0,\ldots,0)_{1\times500}^{\mathsf T}.
\end{equation}
This shift does not modify the features or their internal ordering; it only changes when the corresponding sequence appears in the ordered representation. In this sense, the distinction is purely temporal: the two classes are presented at different positions along the series.

This interpretation also reflects the sequential nature of learning. The tensor $W$ does not receive all samples from the two classes simultaneously. During learning, samples are presented one at a time, and their order therefore becomes part of the representation. The class-dependent shift exploits this temporal degree of freedom to remove the geometric proximity of the embedded samples without modifying their underlying feature information.

\begin{figure}
\begin{center}
\includegraphics[width=7cm]{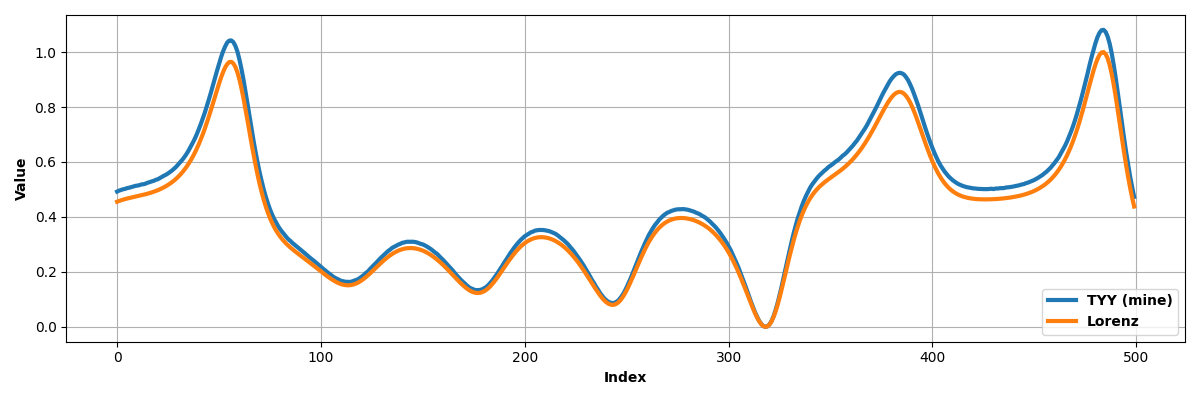}
\includegraphics[width=7cm]{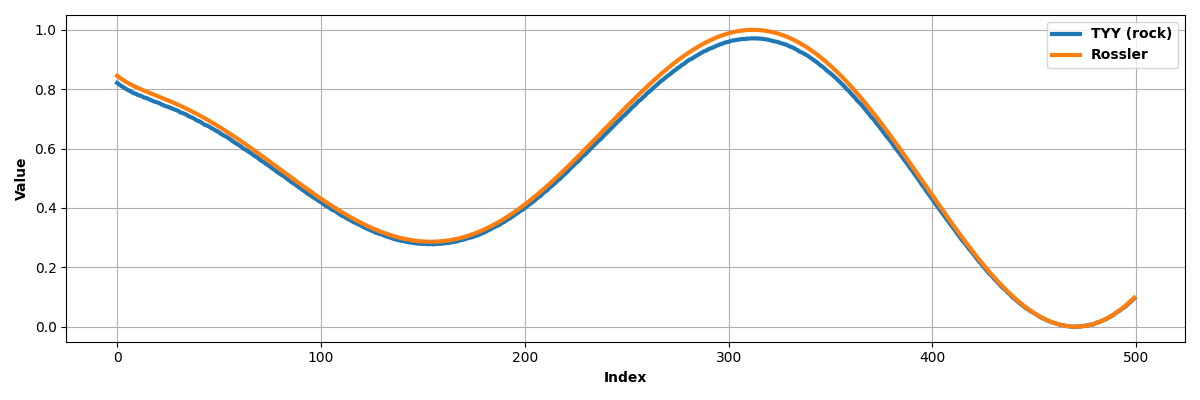}
\caption{Model convergence to the target chaotic attractors after 10 epochs, demonstrating 100\% classification accuracy.}
\label{fig:tensor_fit}
\end{center}
\end{figure}

The tensor $W$ was conceived as a full $N^{3}$ operator, but was implemented in a sparse structured form by exploiting the zero padding of the input vectors. Each sample is embedded in an $N\times 1$ vector such that only a small class-specific index interval contains nonzero values, while all remaining entries are fixed at zero. Therefore, in the quadratic contraction any tensor coefficient associated with a padded index is never activated and makes no contribution to either the output or the gradient. In practice, this means that a large fraction of the nominal $N^3$ tensor is structurally redundant and does not need to be explicitly stored or updated. The computation can thus be restricted to the effective nonzero sub-blocks of $W$, yielding substantial savings in memory usage and reducing the number of arithmetic operations during training. This structured sparsity also speeds convergence, as illustrated in Fig.~\ref{fig:tensor_fit}.

\subsection{Perceptual constancy}
\begin{figure}
\begin{center}
\includegraphics[width=9cm]{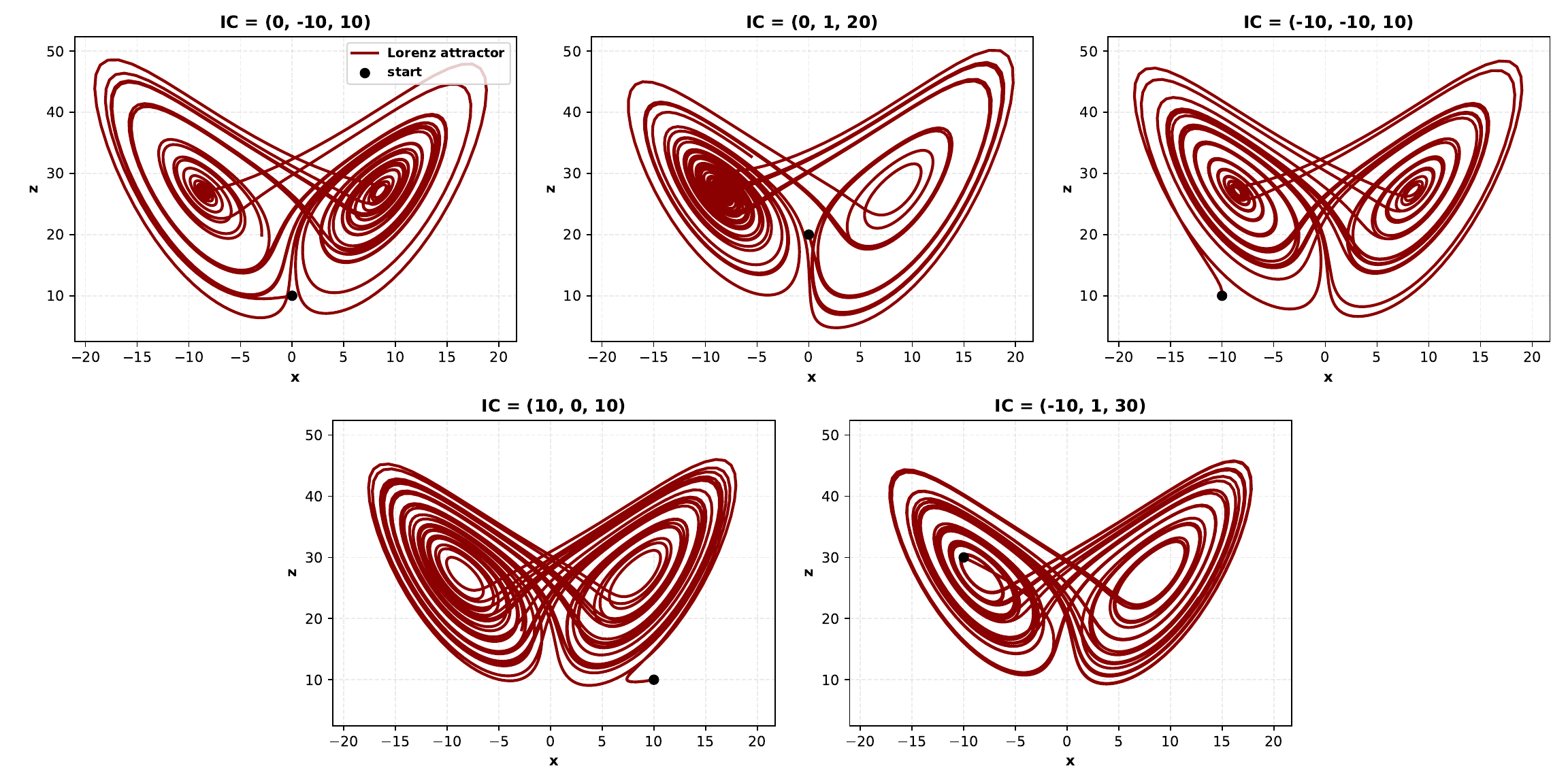}
\includegraphics[width=8cm]{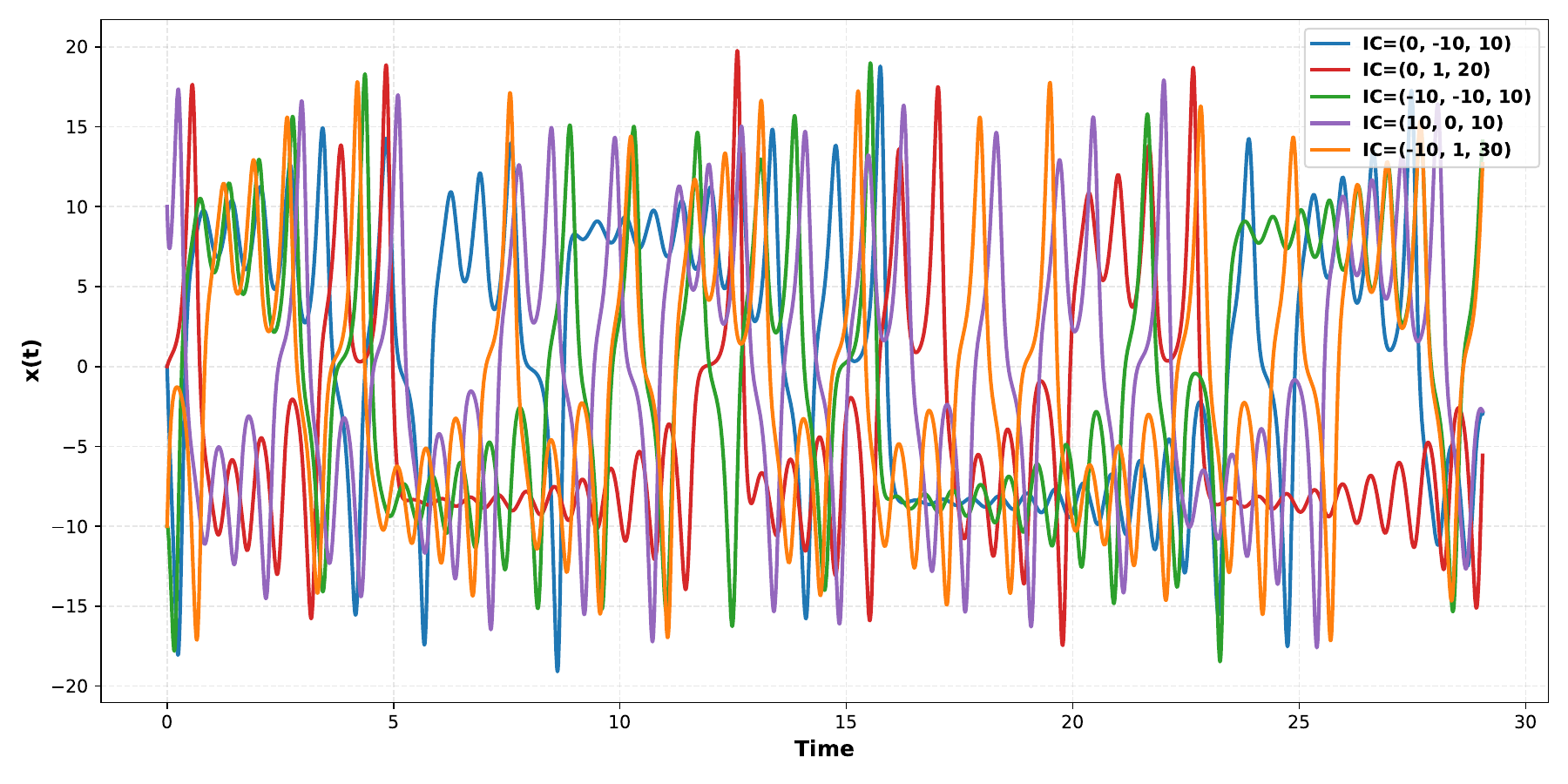}
\caption{5 Lorenz attractors (up) with distinct ($x$-coordinate) time series (down)}
\label{fig:5 Lorenz}
\end{center}
\end{figure}

Objects within the same class are modeled as numerically distinct but topologically equivalent realizations of an attractor. In Fig.\ref{fig:5 Lorenz}, one finds 5 topologically-equivalent Lorenz attractors along with their distinct $x$-coordinates. 

\begin{figure}
\begin{center}
\includegraphics[width=3.5cm]{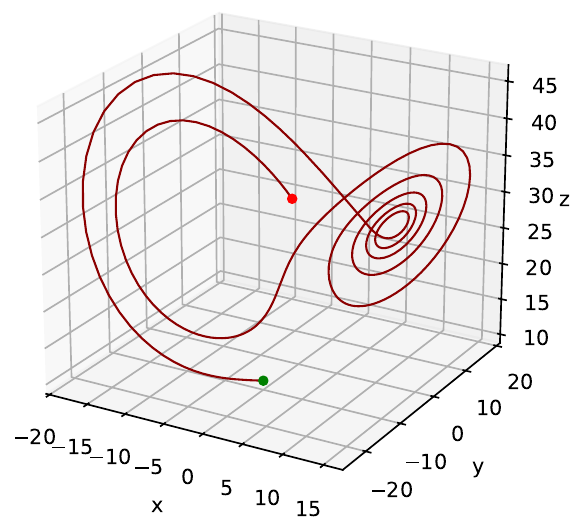}
\includegraphics[width=3.5cm]{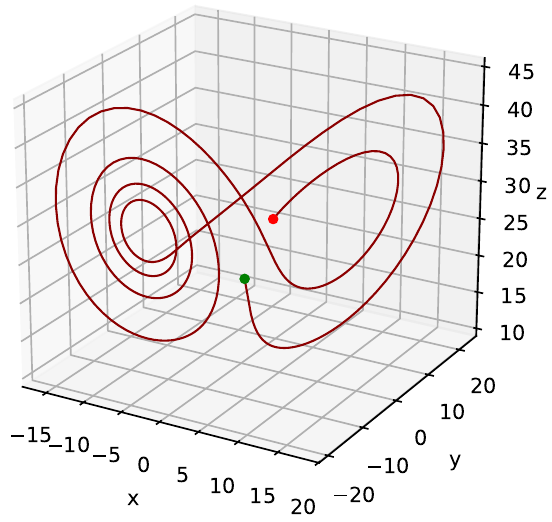}
\includegraphics[width=3.5cm]{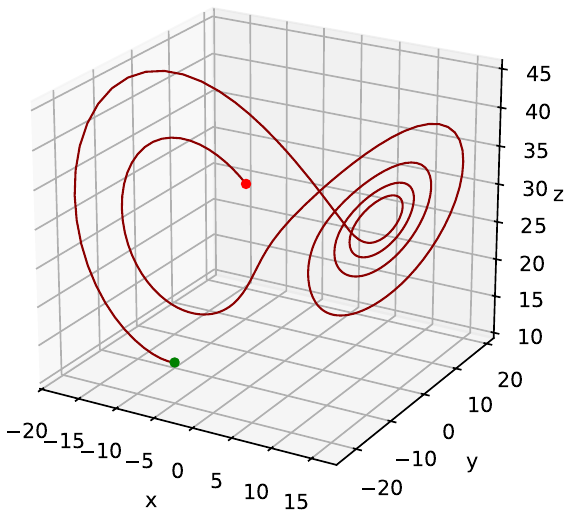}
\includegraphics[width=3.5cm]{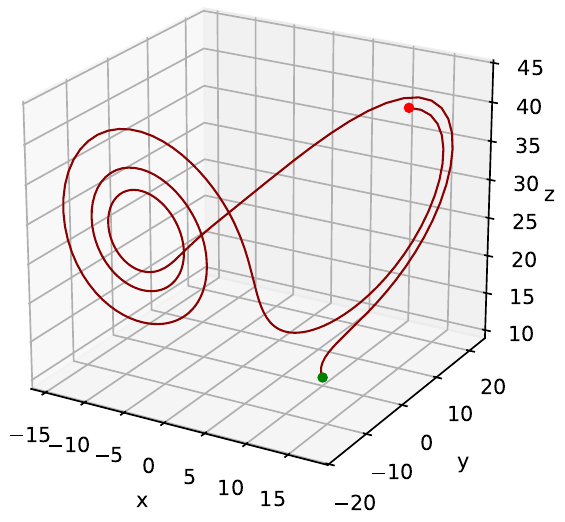}
\includegraphics[width=3.5cm]{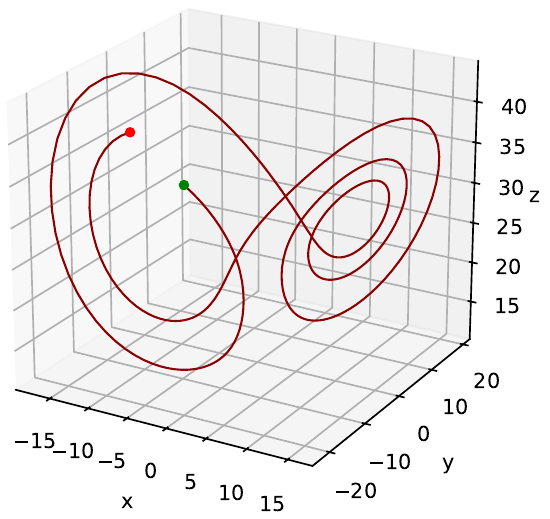}
\caption{
From left to right: the initial points are (0,-10,10),(0,1,20),(-10,-10,10),(10,0,10),(-10,1,30) with their corresponding symbolic dynamics.}
\label{fig:5 Lorenz2}
\end{center}
\end{figure}

By topological equivalence, we mean that all five attractors are characterized by the same linking matrix,
differing, at most, in their diagonal entries\footnote{According to the topological theory of chaotic attractors, only the off-diagonal entries of the linking matrix are topological invariants, whereas the diagonal entries are not uniquely determined.}. The differences in diagonal entries arise from variations in their symbolic dynamics,
\begin{equation*}
\begin{array}{l}
L,LR,LRR,LRRR,LRRRR,\textcolor{blue}{LRRRRR},\cdots\\
R,LR,LLR,LLLR,\textcolor{purple}{LLLLR},\cdots\\
L,LR,LRR,LRRR,\textcolor{red}{LRRRR},\textcolor{blue}{LLRRRR},\cdots\\
R,LR,LLR,LLLR,\textcolor{purple}{LLLRR},\cdots\\
L,LR,LRR,LRRR,\textcolor{red}{LLRRR},\cdots
\end{array}
\end{equation*}
as illustrated in Fig.~\ref{fig:5 Lorenz2}. The significance of the diagonal entries becomes apparent when considering perceptual constancy. 

\begin{figure}
\begin{center}
\includegraphics[width=8cm]{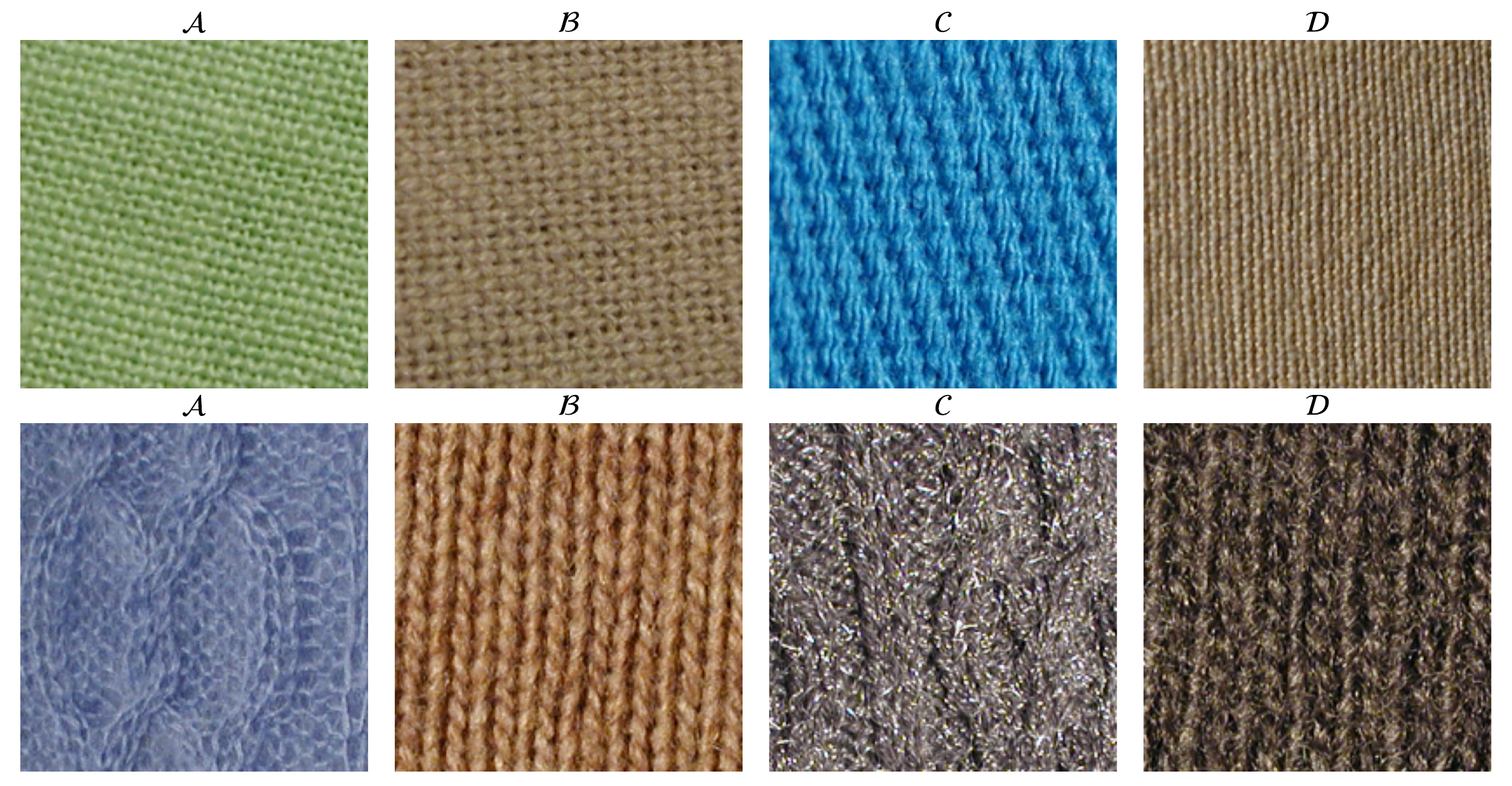}
\caption{Samples–different in color and texture–of two classes, cotton (the first row) and wool (the second row).}
\label{fig:cotton-wool}
\end{center}
\end{figure}

\begin{figure}
\begin{center}
\includegraphics[width=3.5cm]{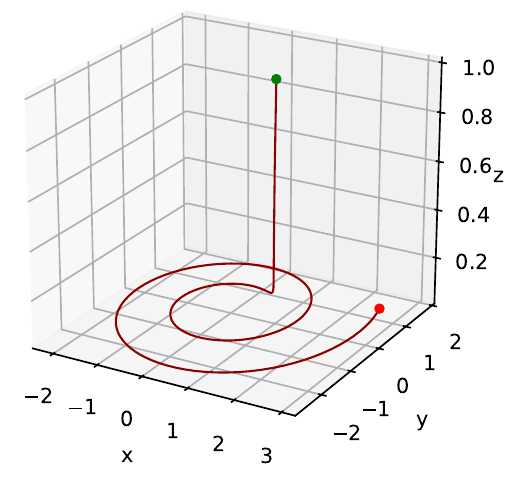}
\includegraphics[width=3.5cm]{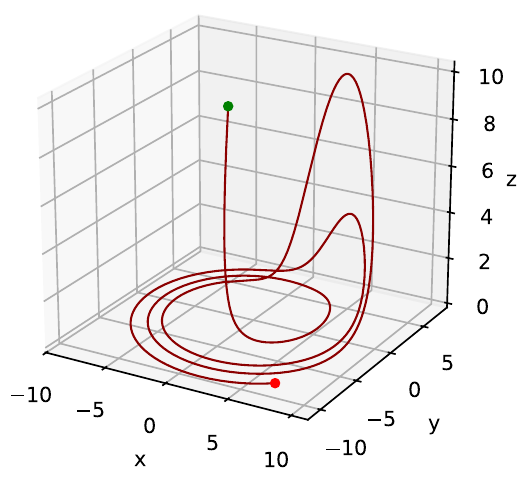}
\includegraphics[width=3.5cm]{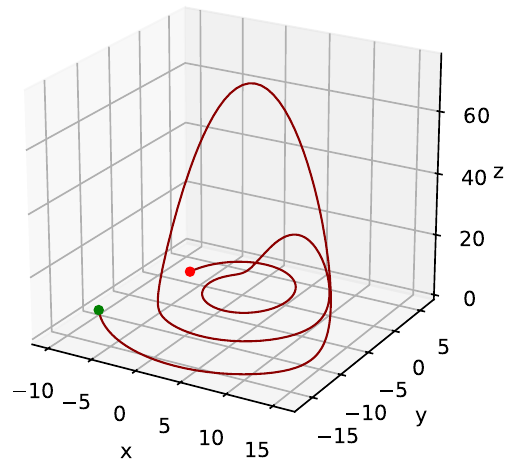}
\includegraphics[width=3.5cm]{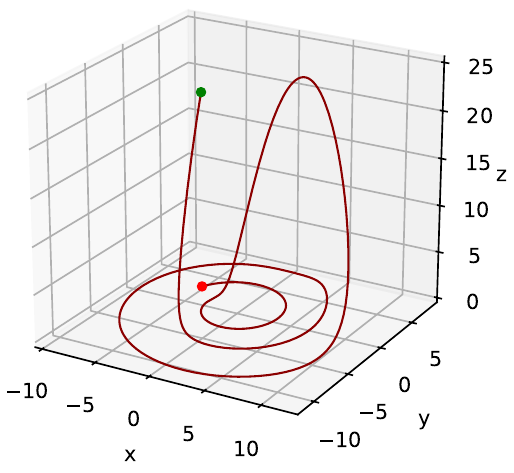}
\caption{R\"{o}ssler attrcators with different symbolic dynamics. From left to right: the initial points are (1,0,1),(1,-5,10),(-10,-10,0),(0,-5,25)}
\label{fig:4 Rossler}
\end{center}
\end{figure}

We consider two classes, namely ``cotton'' and ``wool,'' from KTH-TIPS2-a dataset\cite{Caputo2005} each comprising four sets of images. The images corresponding to the four sample sets are shown in Fig.~\ref{fig:cotton-wool}. For the first class, we assign four Lorenz attractors with different initial conditions, one to each sample set. Similarly, for the second class, we assign four R\"{o}ssler attractors with distinct symbolic dynamics, respectively

\begin{equation}
\begin{split}
&0,00,\textcolor{blue}{000},\cdots\\
&0,00,\textcolor{blue}{001},0010,\textcolor{red}{00101},00101,001010,\cdots\\
&0,01,\textcolor{blue}{010},0101,01010,010100,\cdots\\
&0,00,001,0010,\textcolor{red}{00100},\cdots
\end{split}
\end{equation}

as shown in Fig.~\ref{fig:4 Rossler}.

Given the KTH-TIPS2-a dataset, we seek two global tensors
\begin{equation*}
\begin{split}
W^{\textit{cotton}}=(W^{\textit{c}}_{R},W^{\textit{c}}_{G},W^{\textit{c}}_{B})\\
W^{\textit{wool}}=(W^{\textit{w}}_{R},W^{\textit{w}}_{G},W^{\textit{w}}_{B})
\end{split}
\end{equation*}
that realize the mappings
\begin{equation*}
\begin{split}
W^{\textit{c}}_{R,G,B}:Y^{c}_{R,G,B}\to L_{x,y,z}\\
W^{\textit{w}}_{R,G,B}:Y^{c}_{R,G,B}\to R_{x,y,z}
\end{split}
\end{equation*}
where $L,R$ denote the Lorenz and R\"{o}ssler attractors, respectively. 

\begin{figure}
\begin{center}
\includegraphics[width=8.5cm]{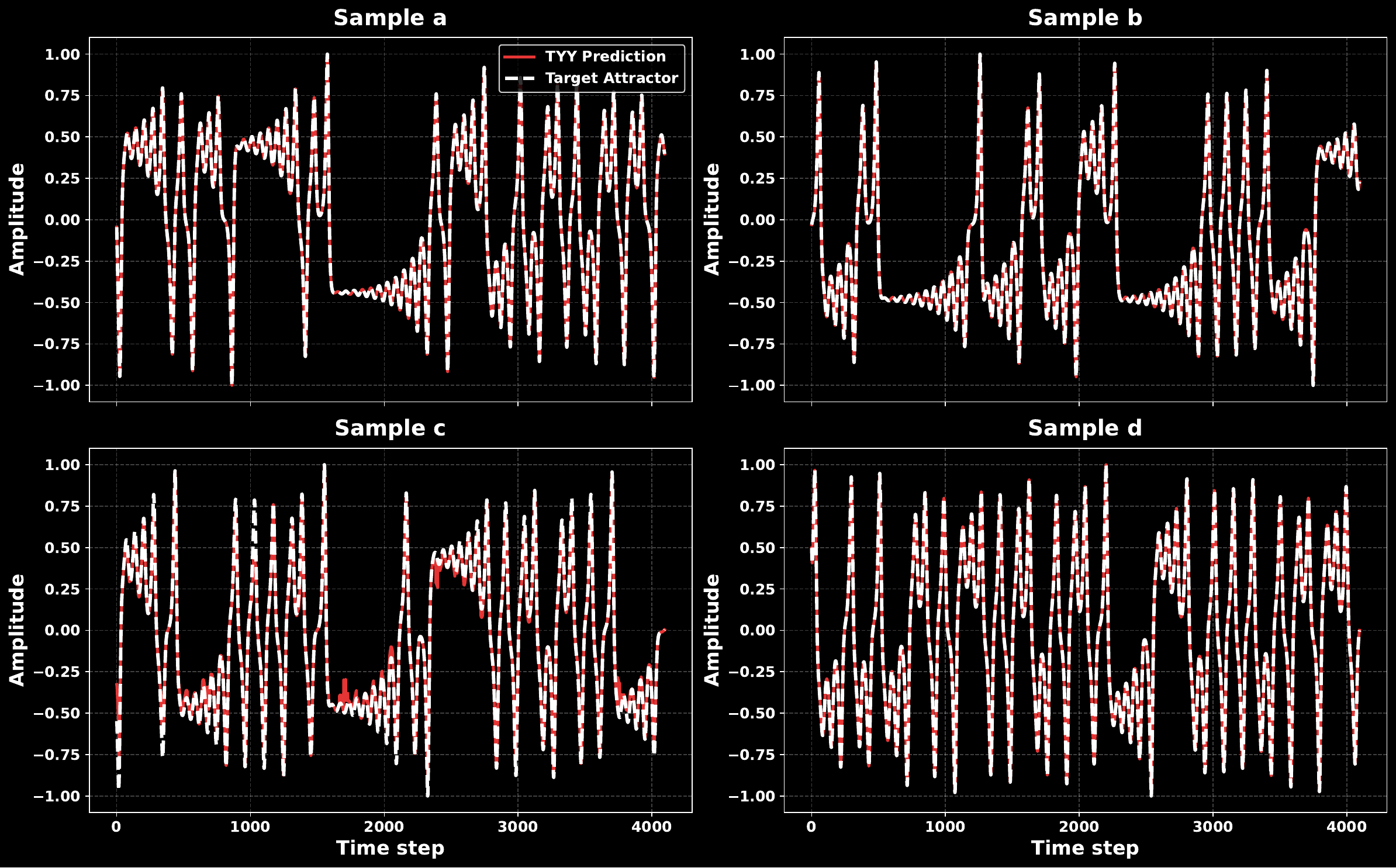}
\caption{The $x$ of Lorenz (dashed) versus \(TYY\) for the red channel of the cotton class after 250 epochs. \(T\) learned from the training set acts on \(Y\) from the test set for samples \(a\), \(b\), \(c\), and \(d\), respectively. Initial standard deviation $=10^{-3}$, learning rate $=5\times 10^{-4}$ and train size $=0.8$.}
\label{fig:R1}
\end{center}
\end{figure}

\begin{figure}
\begin{center}
\includegraphics[width=8.5cm]{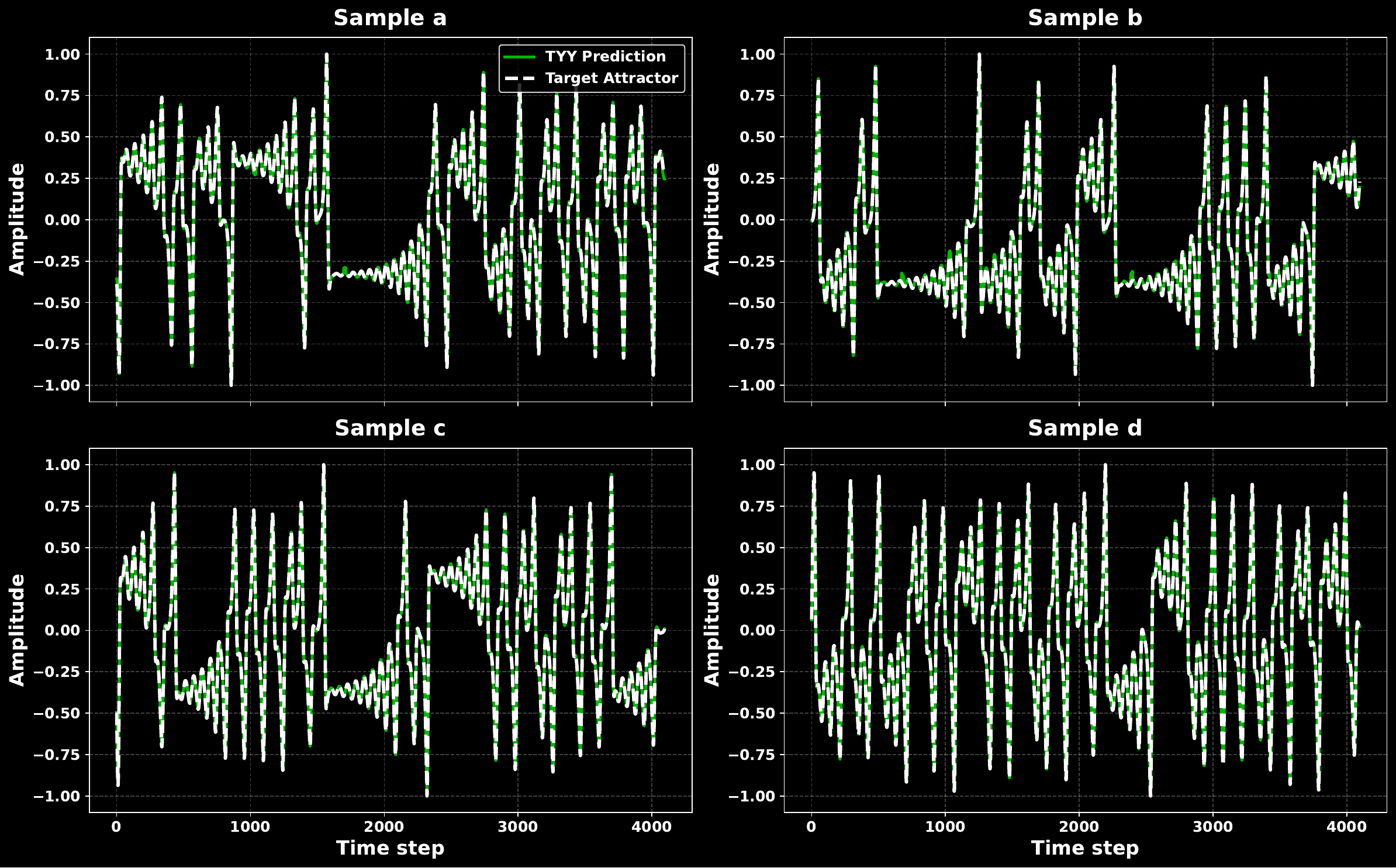}
\caption{The $y$ of Lorenz versus \(TYY\) for the green channel of the cotton class after 15 epochs. Initial standard deviation $=10^{-3}$, learning rate $=5\times 10^{-4}$ and train size $=0.8$.}
\label{fig:R2}
\end{center}
\end{figure}

\begin{figure}
\begin{center}
\includegraphics[width=8.5cm]{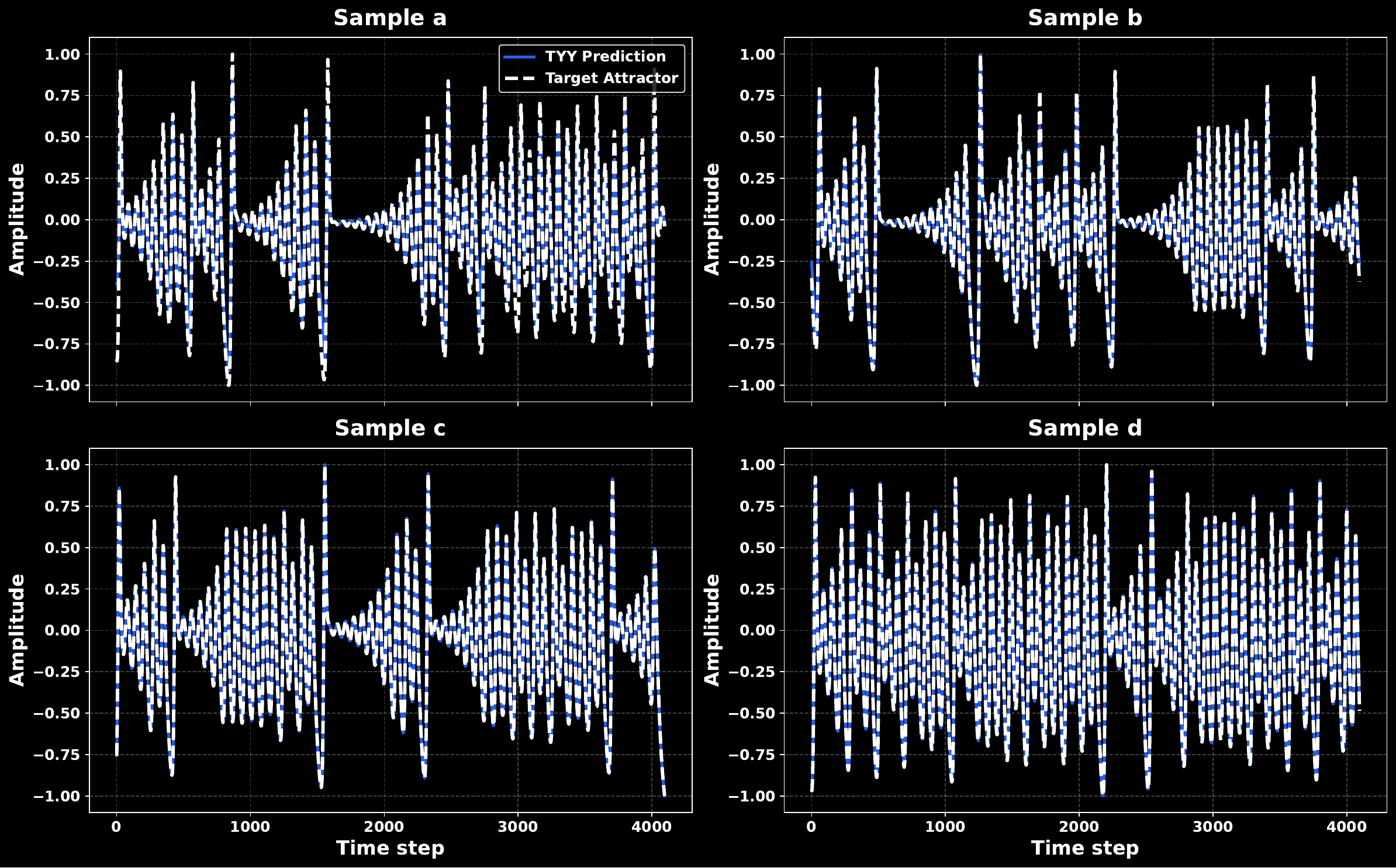}
\caption{The $z$ of Lorenz versus \(TYY\) for the blue channel of the cotton class after 200 epochs. Initial standard deviation $=10^{-3}$, learning rate $=10^{-4}$ and train size $=0.8$.}
\label{fig:R3}
\end{center}
\end{figure}

\begin{figure}
\begin{center}
\includegraphics[width=8.5cm]{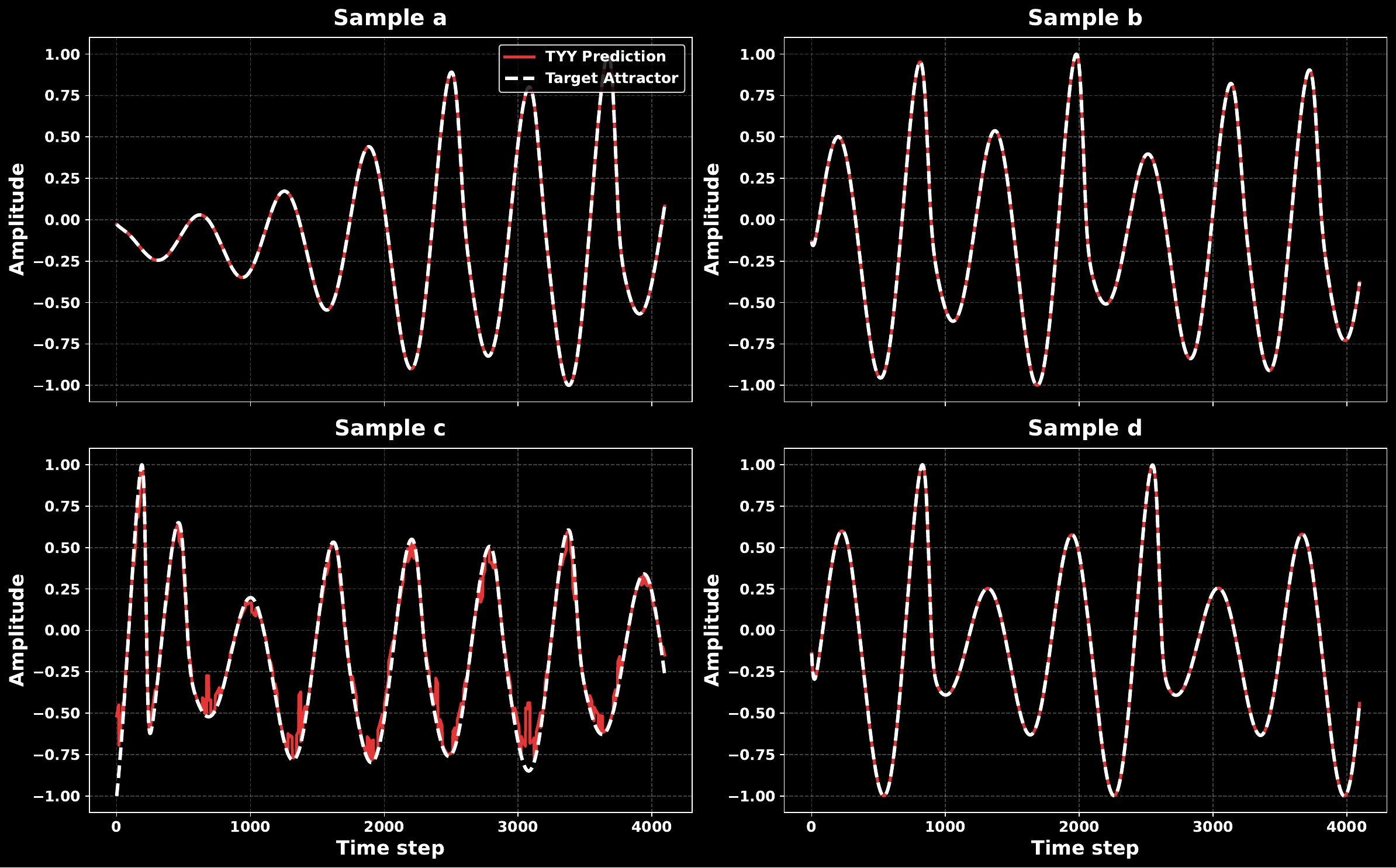}
\caption{The $x$ of R\"{o}ssler versus \(TYY\) for the red channel of the wool class after only 200 epochs. Initial standard deviation $=10^{-3}$, learning rate $=5\times 10^{-4}$ and train size $=0.8$.}
\label{fig:R4}
\end{center}
\end{figure}

\begin{figure}
\begin{center}
\includegraphics[width=8.5cm]{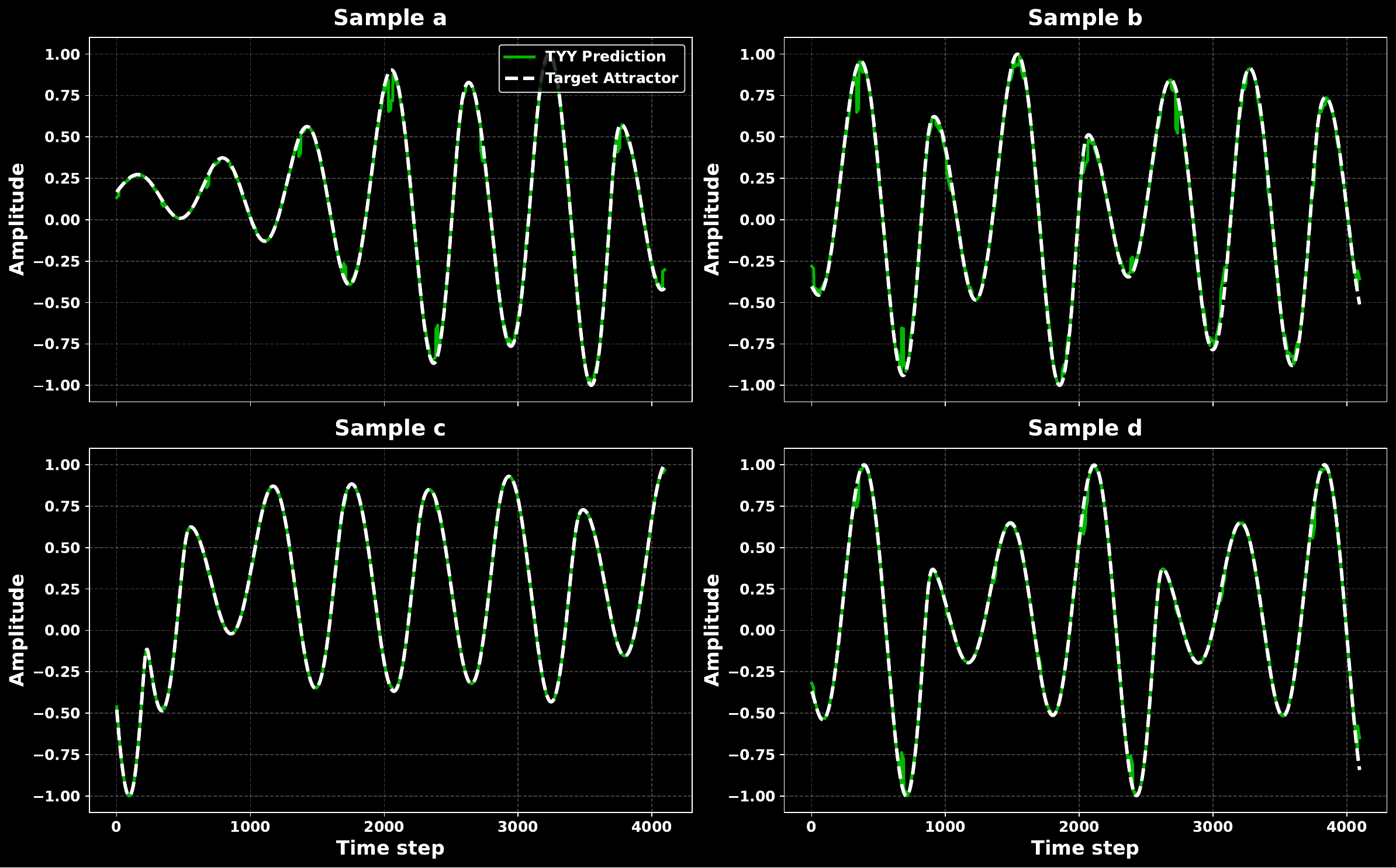}
\caption{The $y$ of R\"{o}ssler versus \(TYY\) for the green channel of the wool class after 50 epochs. Initial standard deviation $=10^{-3}$, learning rate $=2\times10^{-4}$ and train size $=0.8$.}
\label{fig:R5}
\end{center}
\end{figure}

\begin{figure}
\begin{center}
\includegraphics[width=8.5cm]{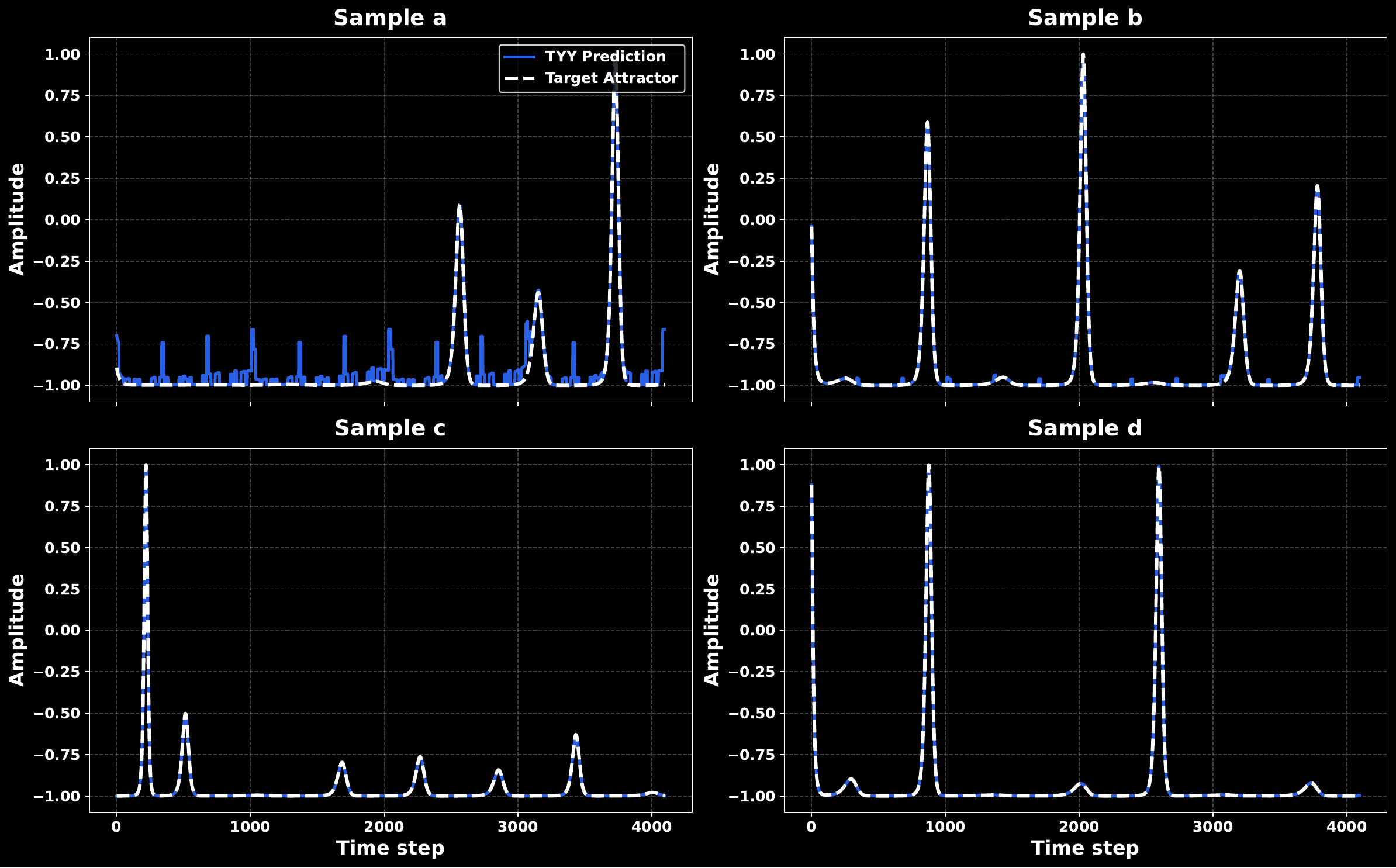}
\caption{The $z$ of R\"{o}ssler versus \(TYY\) for the blue channel of the wool class after 500 epochs. Initial standard deviation $=10^{-3}$, learning rate $=10^{-4}$ and train size $=0.8$.}
\label{fig:R6}
\end{center}
\end{figure}

Like before, we will benefit from vector-series duality the following way: 
\begin{equation}
\begin{aligned}
\mathbf{Y}^{(\mathcal{A})}
&=
\begin{pNiceArray}{c|c|c|c}
\mathbf{S}_{\mathcal{A}}^{(\alpha)} & \boldsymbol{0} & \boldsymbol{0} & \boldsymbol{0}
\end{pNiceArray}^{\mathsf T},
\\
\mathbf{Y}^{(\mathcal{B})}
&=
\begin{pNiceArray}{c|c|c|c}
\mathbf{0} & \mathbf{S}_{\mathcal{B}}^{(\alpha)} & \mathbf{0} & \mathbf{0}
\end{pNiceArray}^{\mathsf T},
\\
\mathbf{Y}^{(\mathcal{C})}
&=
\begin{pNiceArray}{c|c|c|c}
\mathbf{0} & \mathbf{0} & \mathbf{S}_{\mathcal{C}}^{(\alpha)} & \mathbf{0}
\end{pNiceArray}^{\mathsf T},
\\
\mathbf{Y}^{(\mathcal{D})}
&=
\begin{pNiceArray}{c|c|c|c}
\mathbf{0} & \mathbf{0} & \mathbf{0} & \mathbf{S}_{\mathcal{D}}^{(\alpha)}
\end{pNiceArray}^{\mathsf T}
,\quad \mathbf{Y}\in\mathbb{R}^{4096\times1}
\end{aligned}
\end{equation}
$\mathbf{S}\in \mathbb{R}^{1024\times1}$ being the signal (red, green, and blue). Accordingly, $\boldsymbol{J}$s will read 
\begin{equation}
\begin{aligned}
\boldsymbol{J}^{(\mathcal{A})}
&=
\begin{pNiceArray}{c|c|c|c}
\ast & \boldsymbol{0} & \boldsymbol{0} & \boldsymbol{0}
\end{pNiceArray},
\\
\boldsymbol{J}^{(\mathcal{B})}
&=
\begin{pNiceArray}{c|c|c|c}
\boldsymbol{0} & \ast & \boldsymbol{0} & \boldsymbol{0}
\end{pNiceArray},
\\
\boldsymbol{J}^{(\mathcal{C})}
&=
\begin{pNiceArray}{c|c|c|c}
\boldsymbol{0} & \boldsymbol{0} & \ast & \boldsymbol{0}
\end{pNiceArray},
\\
\boldsymbol{J}^{(\mathcal{D})}
&=
\begin{pNiceArray}{c|c|c|c}
\boldsymbol{0} & \boldsymbol{0} & \boldsymbol{0} & \ast
\end{pNiceArray}.
\end{aligned}
\end{equation}
where $\ast$ denotes a nonzero $4096\times1024$ matrix. Under this connectivity scheme, the entire network, comprising 4096 neurons, is divided into four clusters. Within each cluster, the neurons are fully connected and receive the corresponding sample. The one-way connections between clusters extend the attractor and are required for the extraction of a global tensor. $\ast$ follows from Morton ordering, which better preserves spatial locality than row-major raster ordering and is therefore more compatible with local shifts of visual attention. In Figs.\ref{fig:R1}-\ref{fig:R6}, you can find the results for the cotton and wool classes.
 
\section{Conclusion}
In this work, which is not a continuation of previous studies, we propose a learning model based on invariant sets of fractal dimensions. Unlike classical and deep learning models, the weights are represented by a third-order tensor rather than a second-order matrix. The tensor receives an input and produces the corresponding connection matrix, thereby mapping the object to its associated attractor. The connection matrix consequently adapts to the input received by the tensor.

The third-order tensor exploits the vector–series duality proposed by the model by augmenting the dimensionality of the inputs, thereby increasing the length of the chaotic series. This lifting is essential: without it, a global tensor capable of mapping different inputs to their corresponding connection matrices cannot be constructed. Once the dimensionality is lifted, different zero-padding schemes allow the temporal ordering of inputs to be altered. Rather than presenting the inputs to the tensor simultaneously, we place each input at a different position within the lifted vector and present them one at a time. This is consistent with real-life experience, in which objects are encountered and learned sequentially rather than simultaneously. The increased dimensionality also provides sufficient sparsity, facilitating fast fitting and, consequently, fast learning.

The objective of this paper is not to propose a learning model that competes with current classical or deep learning approaches in terms of performance. Rather, we demonstrate a possible application of chaos to real-world learning and explore whether chaotic dynamics can provide a basis for perception and learning.

\section*{Declarations}

\subsection*{Competing Interests}
The author declares that there are no known competing financial or personal
relationships that could have appeared to influence the work reported in this paper.

\subsection*{Data and Code Availability}
The datasets used in this study are publicly available from the respective
sources cited in the manuscript. No new datasets were generated or collected
as part of this study.

The code used to implement the proposed model and reproduce the results is
available from the author upon reasonable request.

\bibliography{REF}
\end{document}